# Navigation-grade interferometric air-core antiresonant fibre optic gyroscope with enhanced thermal stability

Maochun Li[1,†], Shoufei Gao[2,3,5,†], Yizhi Sun[2,3,5,†], Xiaoming Zhao[1,*], Wei Luo[1], Qingbo Hu[2,3], Hao Chen[2,3], Helin Wu[2,3], Fei Hui[1], Yingying Wang[2,3,5*], Miao Yan[1,4], and Wei Ding[2,3,5,6*]

[1]Tianjin Key Laboratory of Quantum Precision Measurement Technology, Tianjin Navigation Instruments Research Institute, Tianjin 300131, China

[2]Guangdong Provincial Key Laboratory of Optical Fibre Sensing and Communication, Institute of Photonics Technology, Jinan University, Guangzhou 510632, China

[3]College of Physics & Optoelectronic Engineering, Jinan University, Guangzhou 510632, China

[4]School of Mechanical Engineering, Nanjing University of Science and Technology, Nanjing 210094, China

[5]Linfiber Technology (Nantong) Co., Ltd. Jiangsu 226010, China

[6]Peng Cheng Laboratory, Shenzhen 518055, China

[†]These authors contributed equally to this work

## Abstract

We present a groundbreaking navigation-grade interferometric air-core fibre optic gyroscope (IFOG) using a quadrupolar-wound coil of four-tube truncated double nested antiresonant nodeless fibre (*t*DNANF). This state-of-the-art *t*DNANF simultaneously achieves low loss, low bend loss, single-spatial-mode operation, and exceptional linear polarization purity over a broad wavelength range. Our 469 m *t*DNANF coil demonstrated a polarization extinction ratio (PER) of ~20 dB when illuminated by an amplified spontaneous emission (ASE) source spanning 1525-1565 nm. Under these conditions, the gyro archives an angular random walk (ARW) of 0.0038 deg h$^{-1/2}$ and a bias-stability (BS) drift over 8500 s of 0.0014 deg h$^{-1}$, marking the first instance of navigation-grade performance in air-core FOGs. Additionally, we validated the low thermal sensitivity of air-core FOGs, with reductions of 9.24/10.68/6.82 compared to that of conventional polarization-maintaining solid-core FOGs of the same size across various temperature ranges. These results represent a significant step towards long-standing promise of high-precision inertial navigation applications with superior environmental adaptability.

## Introduction

The fibre optic gyroscope (FOG), which relies on the Sagnac effect, is one of the most successful optical fibre sensors and serves as the core equipment for inertial navigation, positioning, and attitude determination[1-3]. Due to their high resolution and simple structure, often referred to as the minimum scheme[1], closed-loop interferometric FOGs (IFOGs) have been widely employed in both military and civilian fields, including aviation, aerospace, weapon systems, autonomous vehicles, oil platforms, and well logging, often preferred over ring laser gyroscopes[4,5] and microelectromechanical system gyroscopes[6]. The comprehensive capabilities of the IFOG make it the primary candidate for inertial navigation systems both currently and in the foreseeable future, necessitating further optimization.

The prime performance of a FOG can be analysed from three aspects: short-term noise (e.g., angular random walk, ARW), long-term bias drift (e.g., bias stability/bias instability, BS/BI), and environmental adaptability, as outlined in Fig. 1. Decades of technological efforts to improve all FOG components, including the optical source, modulation strategy, coil winding, and detector, have pushed the IFOG to the pinnacle of these performance metrics[7]. However, the physical attributes of the fibre medium impose fundamental limits on further improvement of IFOGs. In conventional polarization-maintaining silica core fibres (SCFs), several deleterious effects inherently exist for gyro, such as the temperature Shupe effect[8], magnetic field Faraday effect[9], optical nonlinearity Kerr effect[10], Rayleigh backscattering[11], and radiation sensitivity[12]. High-precision IFOGs usually employ kilometer-level fibre coils, where the accumulation of these components amplifies the environmental influence. A possible solution is to utilize a resonant FOG (RFOG) with a shorter fibre length[13]. However, the accuracy of RFOG has only recently reached navigation grade[14] and still markedly lags behind that of IFOGs.

Another promising approach to address environmental issues is to replace the fibre medium of FOGs from silica with air. Exploration in this direction dates back to the year 2006[15]. This change of the fibre medium dramatically reduced errors caused by the Faraday effect (by more than 20 times) and the Kerr effect (by more than 170 times)[16]. Initial research[17] predicted the thermal sensitivity of air-core FOGs to be 3-11 times smaller than that of SCF FOGs, and later experiments revealed a 6.5-fold reduction[18] with photonic bandgap-type hollow core fibres (PBG-HCFs)[19]

In the past decade, another type of HCF, referred to as antiresonant HCFs (ARFs)[20,21], has made significant advances in attenuation reduction[22-24] and effective single-mode operation[25], facilitating the propagation length to reach the kilometer level and mitigating multipath interference, which severely impairs the noise and drift performance of PBG-HCF-based FOGs[15,16,18]. Furthermore, recent experiments have validated that ARFs have better light transmission attributes than PBG-HCFs, such as an almost four orders of magnitude lower nonlinear coefficient[26] and Brillouin gain coefficient[27] than standard single-mode fibres (SSMFs), a lower backscattering coefficient (28 dB below that of SSMF[28]), virtually no radiation-induced attenuation[29], and up to 20-fold reduction in thermal phase sensitivity compared to SSMFs[30]. ARFs, therefore, simultaneously possess advantages in environmental adaptability and modal purity compared to conventional SCFs and PBG-HCFs, respectively, implying their great potential for use in high-performance FOGs.

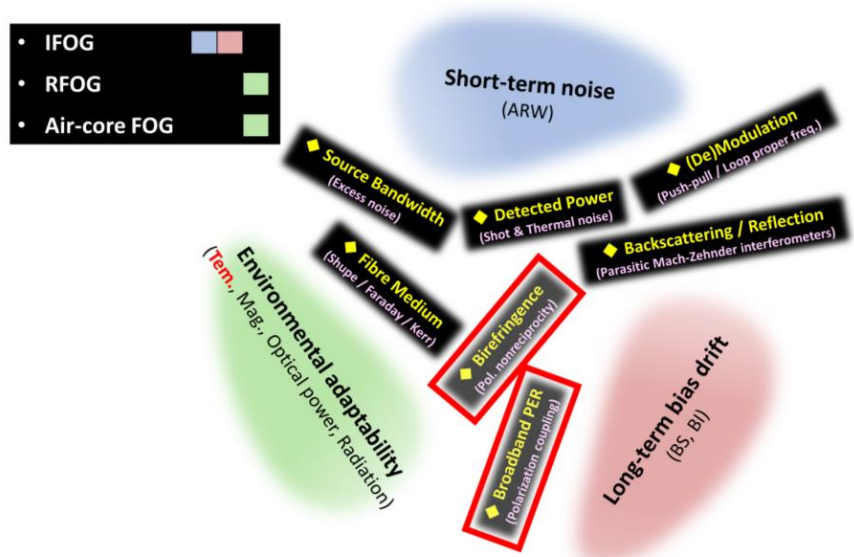

**Fig. 1 | Key metrics (yellow diamonds) of a FOG assessed from three performance aspects (labelled in blue, red, and green, respectively).** While practical IFOGs have almost reached ultimate performances in short-term noise (e.g., ARW) and long-term bias drift (e.g., BI and BS), further improvements in environmental adaptability --- including tolerances to variations of temperature, magnetic field, optical power, and radiation --- are still under intense investigations, with possible solutions being explored by means of RFOGs and air-core FOGs. The primary challenge lies in mitigating all the deleterious effects (described in parentheses and in magenta) concurrently.

However, thus far, ARF-based FOGs have only been tested in RFOG configurations, with implementations of either a 5.6 m Kagome fibre ring[31] or a 136 m nested antiresonant nodeless fibre (NANF) resonator[32]. The observed bias stabilities, albeit greatly superior to those of PBG-HCF-based FOGs, remain at 0.15 deg h$^{-1}$ and 0.05 deg h$^{-1}$, respectively, which are still inferior to the practically usable levels.

The key to unlocking the full potential of ARF-based FOGs may lie in utilizing broadband sources[33], thereby obtaining the advantages of IFOG and maintaining polarization[34]. For the latter, the polarization behaviors in an ARF have two features. On the one hand, ARFs can maintain exceptional polarization purity by overcoming the fundamental anisotropic Rayleigh scattering limit[35] in SCFs. In ARFs, the polarization crosstalk primarily originates from the rough interfaces between air and silica rather than from the air core. The interpolarization power coupling coefficient can be more than three orders of magnitude lower than that in SCFs[36]. On the other hand, state-of-the-art ultralow-loss ARFs are usually non-polarization-maintaining. The principal axis angles of these ARFs have apparent wavelength dependence and are highly sensitive to bending and twisting, meaning that the linear polarization state can only be well preserved within a small wavelength range[36]. As a result, it is vital to balance the broadband light source and linear polarization maintenance in an IFOG of a small-radius coil.

In this work, a low-loss, low-bend-loss four-tube truncated double nested antiresonant nodeless hollow core fibre ($t$DNANF)[37,38] was designed and fabricated. The modest $10^{-6}$ level of birefringence and the principle-axis-angle-holding capability, inherently facilitated by its geometry, led to an approximately 20 dB thermally stable polarization extinction ratio (PER) in a 469 m quadrupolar-wounded fibre coil over the wavelength range of 1525-1565 nm. The combination of all beneficial factors—a quadrupolar-wound coil of sufficient fibre length, simple fibre-to-chip connections, a high spatial mode purity, and a high linear polarization purity—in such an IFOG yields an ARW at a high power of 0.0038 deg h$^{-1/2}$, a BS drift over 100 s of 0.023 deg h$^{-1}$, and a BS drift over 8500 s of 0.0014 deg h$^{-1}$, representing, to the best of our knowledge, the first navigation-grade air-core FOG. Furthermore, a decrease in thermal sensitivity by factors of 9.24/10.68/6.82, compared to that of a conventional polarization-maintaining SCF-FOG with the same coil length and diameter, was validated when our four-tube $t$DNANF-FOG was subjected to different temperature ramps. This exhibited better thermal stability than all the reported PBG-HCF-based FOGs[16,18].

## Results

### Four-tube *t*DNANF structure

Designing an advanced HCF for high-precision FOGs requires a combination of low loss, low bend loss, single modality, and high linear polarization purity. These attributes should be maintained across a broad wavelength range, especially in the context of an IFOG configuration.

To reduce the loss of an ARF, introducing additional layers of glass membranes or nested tubes in the cladding area[37-39] is crucial. Recently, a DNANF has achieved a record loss of <0.11 dB km$^{-1}$ at 1550 nm, surpassing all other optical fibres[24]. Higher-order mode suppression, defined as the ratio of leakage losses of fundamental and higher-order modes, can be enhanced by designing cavities among the cladding tubes to selectively out-couple higher-order modes[25,40]. The DNANF structure is thus ideal for achieving low loss and single-mode operation. For generic ARFs, reducing bend loss involves using a small core with a diameter-to-wavelength ratio less

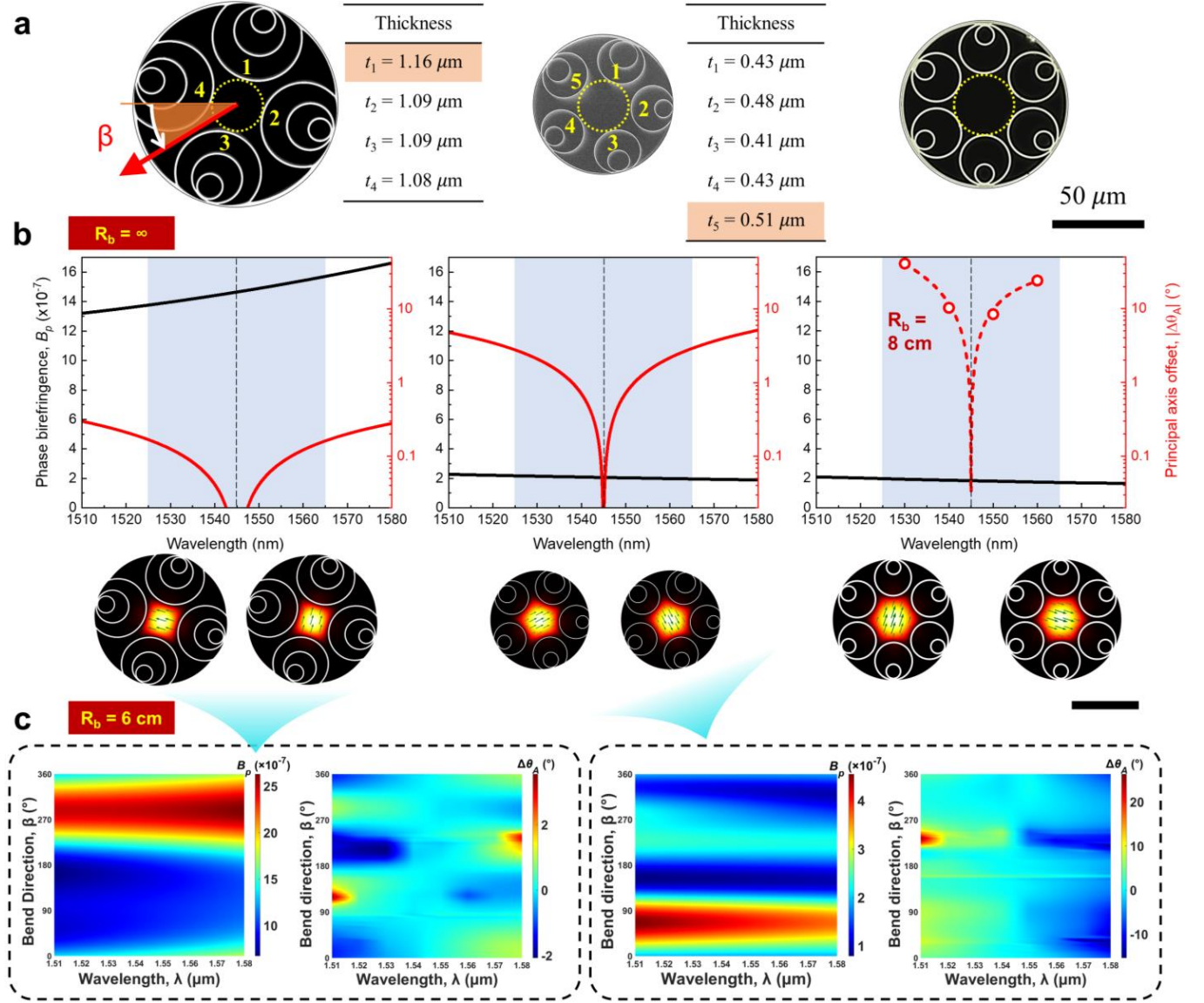

**Fig. 2 | Finite element modelling of polarization properties of three ARFs. a** Scanning electron micrograph cross-sections of the four-tube *t*DNANF and the five-tube NANF fabricated in-house, alongside the six-tube NANF reported in Ref. 36 (referred to as NANF-SI therein). The core diameters of the three ARFs are 28.2 $\mu$m, 27.9 $\mu$m, and 33.4 $\mu$m, respectively. The outer tube thicknesses are 1.08 - 1.16 $\mu$m, 0.41 - 0.51 $\mu$m, and 1.22 $\mu$m, respectively. The gap sizes are measured to be 5.4 - 6.2 $\mu$m, 5.3 - 7.1 $\mu$m, and 1.8 - 3.1 $\mu$m, respectively. **b** Simulated phase birefringence ($B_p$) and principal axis offsets ($\Delta\theta_A$, with respect to that at 1545 nm) with no bend and in the wavelength range of 1510 - 1580 nm. The structural asymmetry residual in realistic ARFs – e.g., the variation of the membrane thickness of the outer tubes, as outlined in panel **a** – gives rise to wavelength-dependent $B_p$ and $\Delta\theta_A$. For the six-tube NANF, the experimental results of $\theta_A$ under a bend radius of 8 cm are acquired from Figure S4 in Ref. 36, and the simulation of $B_p$ utilizes the structural parameters listed in Table S1 therein together with an invariant membrane thickness of 1.22 $\mu$m. For all the three ARFs, the simulated mode-field profiles of the two polarizations at 1545 nm are plotted below with the arrows indicating the vector direction of the transverse electric field. **c** Simulated $B_p$ and $\Delta\theta_A$ of the four-tube *t*DNANF and the five-tube NANF as a function of wavelength ($\lambda$) and fibre bend direction ($\beta$) under a bend radius of 6 cm. $\Delta\theta_A$ are defined as the angle offsets from the principal axis directions at 1545 nm. The red arrow in panel **a** depicts the bend direction from the fibre toward the centre of curvature.

than 20[21,22].

The main challenge in ARF design is maintaining high-purity linear polarization across a wide wavelength range. This can be achieved in a DNANF framework by adopting fourfold rotational symmetry. In ARFs, the intrinsic birefringence is dictated by nonuniformities among the core surrounding membranes and intertube gaps[41,42], with the former being more manageable. A variation in membrane thickness can create inconsistent antiresonant reflecting conditions. An ARF with fourfold rotational symmetry can efficiently transfer this inconsistency to high birefringence, especially when the core diameter is reduced, to increase the mode field overlap with the membranes[43], although this brings about higher loss. More importantly, if the principal axis direction is primarily dictated by a single segment of the core surround, its wavelength dependence can be greatly mitigated.

We designed and fabricated a four-tube $t$DNANF with a core diameter of 28.2 μm to facilitate low bend loss and high birefringence. The purpose of partly truncating the four outer tubes[44] is to shrink the void regions behind the intertube gaps since these cladding regions may create phase matches with the core fundamental mode. The membrane thicknesses of the outer tubes are measured to be 1.08 - 1.16 μm, corresponding to a normalized frequency (defined as $2t\sqrt{n^2-1}/\lambda$) of ~1.5 at 1550 nm. Here, $t$ is the membrane thickness, $n$ is the refractive index of glass, and $\lambda$ is the wavelength. A normalized frequency close to a half-integer indicates operation in the middle of an antiresonant transmission band, favouring low loss and wavelength independence.

We numerically calculated three realistic ARFs: one four-tube $t$DNANF, one five-tube NANF (both fabricated in-house), and one six-tube NANF reported in Ref. 36. Scanning electron microscopy (SEM) images and structural parameters are presented in Fig. 2a and its caption. All the ARFs have nearly the same core diameters and small gap sizes. For the six-tube NANF (referred to as NANF-SI in Ref. 36), no membrane thickness variation is provided. Using finite element method simulation (see Methods), we obtained birefringence and relative principal axis angles in the 1510-1580 nm range (Fig. 2b). Without bending, the four-tube $t$DNANF exhibits birefringence almost an order of magnitude greater than that of the five-tube and six-tube NANFs. Additionally, the principal axis angle variation of the four-tube $t$DNANF (< 0.4°) is much smaller than that of the five-tube NANF (< 6°) and the six-tube NANF (~ 70°, experimental data) across 1525-1565 nm, which is the wavelength region of our IFOG light source (see Subsection 3).

For FOG applications, it is crucial to consider the impacts of fibre bending on the birefringence and principal axis angle. In the simulation shown in Fig. 2c, with a bend radius of 6 cm (consistent with our IFOG, as detailed in Subsection 3), the birefringence of the four-tube $t$DNANF remains an order of magnitude greater than that of the five-tube NANF. Despite the discontinuous rotational symmetry in fibre cross-sections affecting the polarization properties differently when bent at various orientations, the simulation (Fig. 2c) indicates that the principal axis angle offset of the four-tube $t$DNANF remains less than 2° across the 1525-1565 nm range. This may render a high PER when broadband linearly polarized light is launched into the fibre coil (see Methods and Supplementary Material S1.3).

**Fibre optical characterization**
To validate the analysis of the previous subsection, we characterized the optical properties of propagation loss, macrobend loss, spatial mode purity, the Shupe constant, and linear polarization purity of our four-tube $t$DNANF. The results are shown in Fig. 3.

Figure 3a shows the loss spectrum measured using the cut-back method from 510 m to 10 m, with the fibre wound on a drum with a radius ($R_b$) of 16 cm. This revealed an average propagation attenuation of 0.38 dB km$^{-1}$ within the wavelength range of 1525-1565 nm. To measure the bend loss, a section of the fibre was unwound from the drum and bent to a radius of $R_b$ = 6 cm for 100 turns. The transmission spectra recorded before and after bending indicate macrobend loss, shown by the red line in Fig. 3a, with an average value of ~ 4.7 dB km$^{-1}$. Additionally, Fig. 3b shows the real-time monitored loss during quadrupolar fibre winding (see Methods and Supplementary Material S3.1). A 40 nm bandwidth amplified spontaneous emission (ASE) source and a photodetector were assembled on and rotated with the two supply trays in our fibre winding setup. An overall extra loss of 1.51 dB was detected when the 469 m fibre length was wound with 36 layers and 30 turns, with a 6 cm inner radius and a 7.6 cm outer radius. The linear fit in Fig. 3b indicates an additional loss of 3.3 dB km$^{-1}$, which is consistent with the measured result of macrobend loss at $R_b$ = 6 cm.

To assess the spatial mode purity, a spatially and spectrally resolved imaging (S$^2$-imaging) method[45] was employed for the $t$DNANF coil at 1550 nm (see Methods and Supplementary Material S1.1). As shown in Fig. 3c, no LP$_{11}$ or other higher-order modes are discernible above the noise floor (~ -70 dB) at the output of the fibre coil. Resonant higher-order mode filtering[25], especially under bending state, facilitates such a high level of single-modedness, thus preventing intermodal interference in FOG operation.

To quantify the Shupe constant, which is defined as the relative phase change with temperature, $S = 1/\varphi \cdot d\varphi/dT$, where $\varphi$ is the phase accumulated along the fibre and $T$ is the temperature, an all-fibre Mach–Zehnder interferometer was utilized. In our setup, a 10 m four-tube $t$DNANF coiled with an $R_b$ of 6 cm served as the signal arm and was placed inside a thermal chamber (see Methods and Supplementary Material S1.2). As shown in Fig. 3d, the linear fit of the phase variation at 1550 nm reveals an $S$ = 0.52 ppm °C$^{-1}$, which is 14.4 times smaller than that of the control polarization-maintaining SCF (PMF, $S$ = 7.5 ppm °C$^{-1}$). Our measurement agrees well with the result in Ref. 30, where for a generic NANF, the authors claim that the thermal expansion of fused silica, as a major effect, contributes $S \approx 0.55$ ppm °C$^{-1}$ at room temperature. With a decreasing Shupe constant, the rotation rate error of our four-tube *t*DNANF gyro caused by environmental temperature fluctuations (proportional to $n_{\text{eff}}^2 \times S$ [18]) is expected to be ~ 30 times smaller than that of an SCF gyro. However, it must be noted that in a real fibre coil, the influences of quadrupolar winding and glue cannot be ignored. The Shupe constant of our fibre coil may be higher than 0.52 ppm °C$^{-1}$.

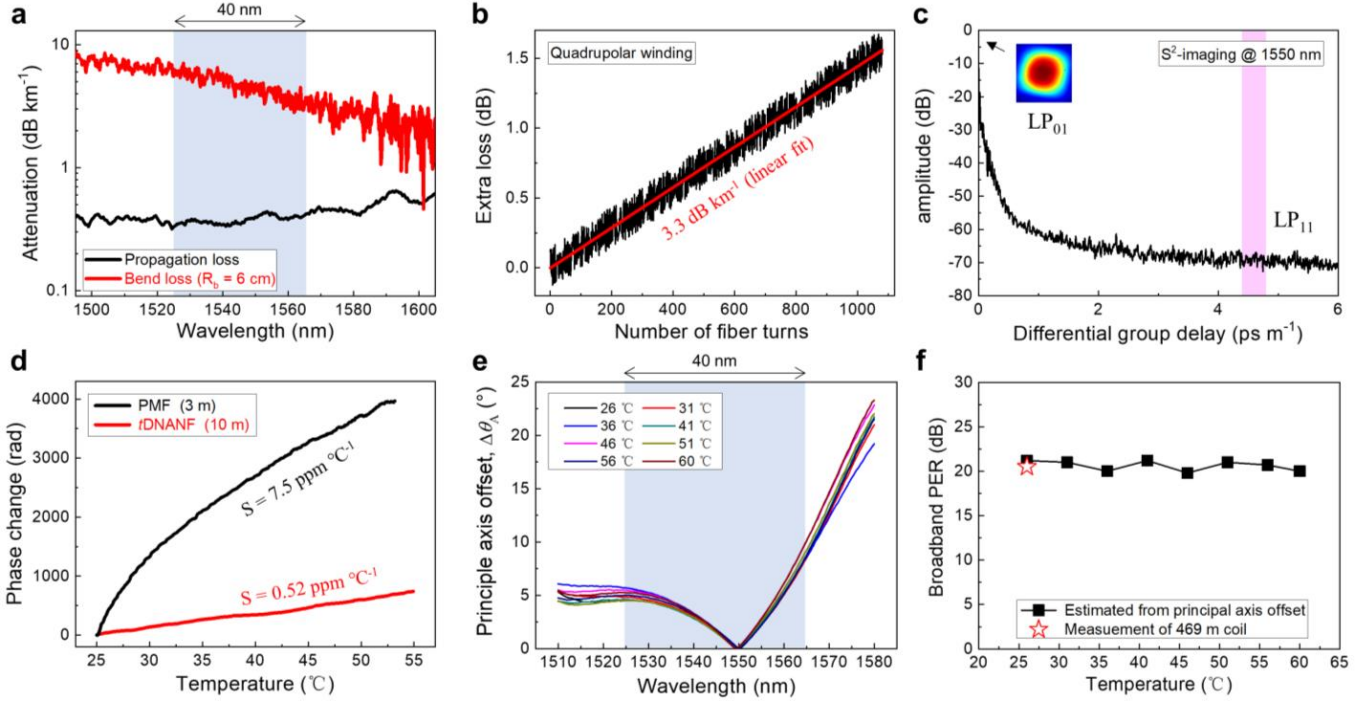

**Fig. 3 | Optical property characterizations of the four-tube *t*DNANF. a** Measured spectral propagation loss and macrobend loss under $R_b$ = 6 cm. **b** Extra loss real-time recorded in a quadrupolar winding. The coil length is 469 m, an ASE source of 40 nm bandwidth is implemented, and the innermost and the outermost radii of the fibre coil are 6 cm and 7.6 cm, respectively. **c** Fourier transform of the optical transmission spectrum of the 469 m quadrupolar-wound coil at 1550 nm measured by $S^2$-imaging method. **d** Measured phase drifts over a 3 m PMF and a 10 m four-tube *t*DNANF at 1550 nm with the temperature increasing from 25 to 55 °C. **e** Measured principal axis offsets relative to the orientations at which minimum powers are detected in a crossed-polarizer setup at various temperatures. A 10 m four-tube *t*DNANF is measured in a 6 cm radius loop configuration. **f** Estimated extinction ratio between the two polarizations of the total integrated power from 1525 to 1565 nm based on the measured principal axis offset in **e** (black squares, see Methods), along with measured PER of the 469 m coil by using an ASE source of 10 nm bandwidth (red star).

A crucial metric of the IFOG configuration for evaluating the polarization-maintaining capability is the broadband PER, which measures how well a fibre (coil) preserves the linear polarization state of transmitted light over a certain wavelength range. The ultimate limit of broadband PER in an ARF is imposed by the wavelength dependence of the principal axis angle[36], which causes the inevitable excitation of the two polarization states by a broadband linearly polarized light source (see details in Supplementary Material S2). Based on the spectrally resolved Jones matrix model in Ref. 36, the orientations of the principal axes at various wavelengths can be determined using a broadband cross-polarizer measurement setup (see Methods and Supplementary Material S2.1). Limited by the resolution of our optical spectrum analyser (AQ6370D, Yokogawa), we measured a 10 m segment of four-tube *t*DNANF. As depicted in Fig. 3e, the offset of the principal axis angle ($\theta_A$) varies by approximately 10° within the wavelength range of 1525-1556 nm. This variation may be attributed to the structural asymmetry discussed in the previous subsection and the fibre bending with $R_b$ = 6 cm. Nevertheless, when the temperature changes from 26 °C to 60 °C, no pronounced variation in the principal axis angle spectrum is observed, aligning well with Ref. 36. The small temperature sensitivity of $\theta_A$ may be due to changes in the microbending conditions, as the Young's modulus of our acrylate coating, which has a glass transition temperature of ~ 40 °C, changes dramatically within this temperature range.

Based on the measured principal axis offset, we can gain insights into the broadband PER of the 469 m *t*DNANF coil with an $R_b$ of 6 cm within the range of 1525–1565 nm (see Methods and Supplementary Material S2.3). As shown by the black squares in Fig. 3f, the estimated broadband PER remains at approximately 20 dB in the varying temperature range. Additionally, the PER of the four-tube *t*DNANF coil was checked using a 10 nm bandwidth ASE source immediately after fibre winding. The result of 20.5 dB (red star in Fig. 3f) closely agrees with the estimation, confirming the high quality of the quadrupolar winding. Note that in Ref. 46, a PER of greater than 30 dB after a 467 m PBG-HCF with no bending was only maintained over a narrow wavelength range of 1530.0-1533.7 nm.

**Gyro test**

The setup of our four-tube *t*DNANF gyro is illustrated in Fig. 4a. The ASE light source based on a 3.2-m erbium-doped fibre has an output power of 8.5 mW. A 1550-nm filter reflector and a gain flat filter are connected to the two ends to provide double-pass amplification and spectrum shaping, respectively, resulting in a bandwidth of ∼40 nm centered at 1544 nm, as shown in Inset 1. Additionally, a fibre-optic isolator ensures monodirectional operation. A polarization-maintaining fibre coupler directs the light emitted from the source to a multifunction integrated optics chip (MIOC), which is fabricated on a lithium niobate substrate. The MIOC consists of a polarizer with a very high extinction ratio (∼70 dB), a Y junction, and two push-pull electro-optic phase modulators for dynamic biasing. The outgoing ports of the Y junction are connected to the 469-m four-tube *t*DNANF coil via fibre-to-chip direct coupling (see Methods and Supplementary Material S3.2). The light returning through the MIOC is directed by the same fibre coupler to a photodetector (PD), whose output currents offer rotation rate information as well as feedback to control the square-wave biasing modulation-demodulation via an electronics package. The modulation depth is set at the maximum sensitivity point ($\pi/2$). The measured eigenfrequency of 320.47 kHz confirms that the effective refractive index of the four-tube *t*DNANF's fundamental mode is ∼1.0, as expected since the fibre core mode travels mostly in air.

A microscopic magnification of the fibre-to-chip coupling assembly is shown in Inset 2 of Fig. 4a. It consists of two sets of microlenses and one reflective microprism. The microlens pairs function as beam collimators and mode field diameter adapters. The reflective microprism, placed between the two microlenses, ensures that the collimated beam from one outgoing port of the MIOC is reflected twice and is set far from the parallel beam emitted from the other port. This separation serves two purposes: (1) preventing crosstalk between the two light beams from the MIOC and (2) providing ample space for handling the two pigtails of the fibre coil, whose transverse sizes amount to ~1 mm after being assembled with microlenses (as shown in inset 3 of Fig. 4a). The imbalance loss between the two pigtails (~0.2 dB) caused by the reflective microprism has a negligible impact on the gyro drift due to the extremely low Kerr nonlinearity in air-core fibres. The output facets of the MIOC and microlenses are 10°- and 8°-cleaved, respectively, to eliminate back-reflection, while the four-tube $t$DNANF is flat-cleaved. During fibre-to-chip assembly, we rotated one end of the fibre coil to maximize the polarization extinction ratio at the other end and then rotated the other end of the fibre coil to maximize the power at the PD, ensuring that the linearly polarized light exiting from the MIOC aligned with the principal axis of the four-tube $t$DNANF. Subsequently, without modulation-demodulation, the round-trip loss from the ASE source to the PD was measured to be 22.4 dB, and the power at the PD was 48.6 μW.

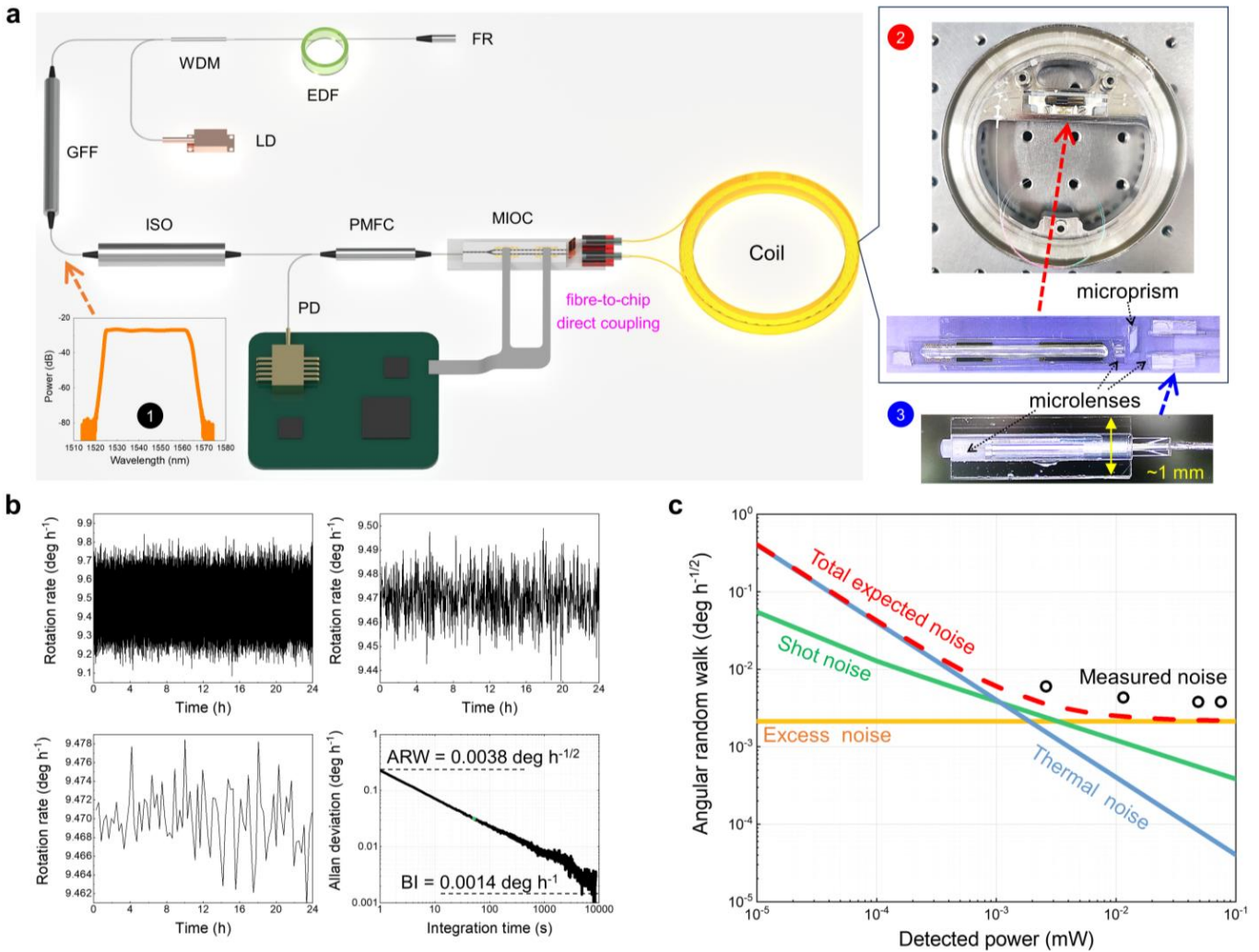

**Fig. 4 | Static performances of the four-tube *t*DNANF gyro. a** Configuration of the IFOG. EDF, erbium-doped fibre; LD, 980 nm laser diode; WDM, 980/1550 nm wavelength-division multiplexer; FR, 1550 nm filter reflector; GFF, gain flattening filter; ISO, fibre-optic isolator; PMFC, polarization-maintaining fibre coupler; MIOC, multifunction integrated-optics chip; PD, photodetector. Inset 1, output spectrum of the ASE source. Inset 2, photographs of the fibre coil and the MIOC with the two outgoing ports free-space coupled to the fibre coil. Inset 3, photograph of a fibre end assembled with a microlens. **b** Static rotation rate readout from the gyro, displaying results over average time windows of 1 s, 100 s, and 1000 s, along with the Allan deviation of the data. **c** Measured ARW noise versus detected power for the gyro, compared with theoretically calculated noise of individual sources.

To assess the operation of the gyro, our four-tube $t$DNANF gyro was mounted on a stable pier with the coil axis perpendicular to the horizontal plane (see Supplementary Material S3.3 and S3.4). The 24-hour response of the gyro to the Earth's rotation at room temperature (25 °C) is displayed in Fig. 4b, which shows the results over average time windows of 1 s, 100 s, and 1000 s, accompanied by the Allan deviation of the data. The measured rotation rate agrees well with the theoretical value of ~ 9.47 deg h$^{-1}$, which is a fraction of the Earth's rotation rate, given that the latitude of the test site is 39°10'.

An ARW of 0.0038 deg h$^{-1/2}$ and a BI of 0.0014 deg h$^{-1}$ can be determined from the Allan deviation curve, demonstrating that the resolution of our four-tube $t$DNANF gyro reached navigation grade. The BI of the gyro is 35.7 times better than that reported in previous NANF-based RFOG[32]. The outstanding bias drift performance can be mainly attributed to the single-mode, low-loss light transmission in the coil (overall loss <4 dB) and the wide spectrum band (FWHM ~40 nm) of the light source. The polarization-sustaining capability of our four-tube $t$DNANF coil, with a measured PER of 20.5 dB, as described in Fig. 3f, also contributes to the low gyro drift.

The noise of our four-tube $t$DNANF gyro mainly arises from three effects. As plotted against the detected power in Fig. 4c, these contributions are excess noise from the ASE source (with an ARW proportional to $P^0$), thermal noise from the PD ($P^{-1}$), and shot noise ($P^{-1/2}$). Each noise contribution is calculated from specifications. The expected total noise (the red dotted curve in Fig. 4c) was obtained by adding up these statistically independent sources of noise quadratically. Thermal noise is expected to dominate at detected powers below 1 μW, and excess noise above 3 μW. The total noise was experimentally verified by varying the output power of the ASE source. These ARW measurements (black circles in Fig. 4c) show very close agreement with the expected total noise.

**Thermal sensitivity**

According to the Shupe effect[8], a thermally induced phase shift between the two counterpropagating beams, which is indistinguishable from the rotation-induced phase shift, occurs in the fibre coil of an IFOG. The generated rotation rate shift (error) $\Omega_E$ can be expressed as[18]

$$\Omega_E = \frac{n^2 SL}{D} \int_0^1 (1-2z')\dot{T}(z')dz', \quad (1)$$

where $n$ denotes the effective index of the core mode (~0.999 for our four-tube $t$DNANF), $S$ is the Shupe constant of the coil, $L$ is the total length of the coil, $D$ is the diameter, $\dot{T}(z)$ is the time derivative of temperature in a fibre element of length $dz$ located at a distance $z$ from one end of the coil, and $z' = z/L$.

The four-tube $t$DNANF gyro was stably placed on a vibration isolation base inside a temperature chamber to ascertain its response to transient temperature variation (see Supplementary Material S3.5). The left panel of Fig. 5a shows the temperature of the chamber, the time derivative of temperature, and the rotation rate shift of the gyro under a nominal temperature change rate of 1 °C per minute within a temperature range of -40 to 60 °C. The temperature was tracked by a thermocouple attached to the supporting structure of the coil. The initial surface temperature of the coil was set at 20 °C, then lowered to -40 °C at the specified rate and maintained for 2 hours before rising to 60 °C at the same rate and holding for another 2 hours. During cooling, the coil temperature decreased steadily with a nonlinear ramping rate, depending on the temperature control accuracy of the chamber. The thermally induced rotation rate shift increased at the beginning of the temperature change and then decreased at the end. During heating, the rotation rate shift had a similar trajectory but with the opposite trend. This characteristic is dictated by both the heat conduction inside the fibre coil and the compensation mechanism to the thermally induced error by symmetrical quadrupole winding. The heat flow is gradually transferred from the outermost to the internal layers, yielding a series of thermally induced phase shifts. Despite partial cancellation of these shifts by means of quadrupolar winding, the overall phase shift continues to accumulate as the outer layers heat faster than the inner layers, resulting in a shift in the rotation rate, which gradually vanishes at the end of the temperature change. Theoretically, the thermally induced rotation rate shift is proportional to the temperature–time derivative. The measured rotation rate shift aligns well with this, verifying the gyro's thermal response.

The evolutions of the four-tube $t$DNANF gyro rotation rate shift at various temperature change rates of 0.2, 0.5, 1, 2, and 5 °C per minute are shown in the right panel of Fig. 5a. The maximum rotation rate shifts were 0.0353 deg h$^{-1}$ (at 5 °C per minute), 0.0162 deg h$^{-1}$ (at 2 °C per minute), 0.0079 deg h$^{-1}$ (at 1 °C per minute), 0.0051 deg h$^{-1}$ (at 0.5 °C per minute), and 0.0013 deg h$^{-1}$ (at 0.2 °C per minute), respectively, confirming that the maximum rotation rate shift increases with the temperature change rate. Similar experiments were conducted within the temperature ranges of 20 to 60 °C and -40 to 20 °C, revealing similar results (Figures 5b and 5c).

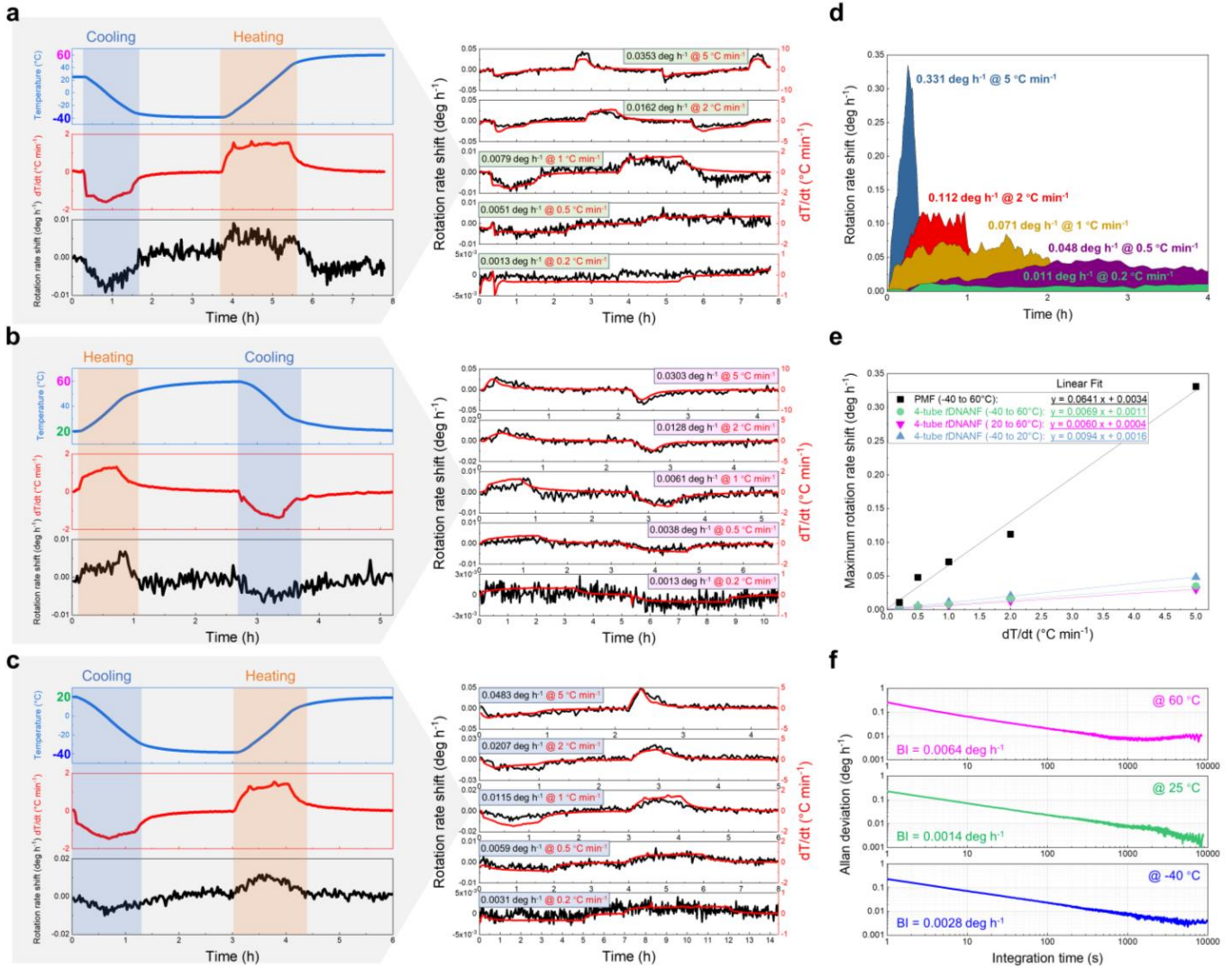

**Fig. 5 | Thermal stability of the four-tube *t*DNANF gyro compared to a conventional polarization-maintaining FOG. a** Surface temperature of the four-tube *t*DNANF coil, temperature time derivatives, and rotation rate shifts of the four-tube *t*DNANF gyro under various temperature change rates of 0.2, 0.5, 1, 2, and 5 °C per minute within a temperature range of -40 to 60 °C. **b,c** Similar to **a**, but for the temperature intervals of 20 to 60 °C and -40 to 20 °C, respectively. **d** Temperature variation measurements for a gyro equipped with a conventional quadrupolar-wound PMF coil (of the same length) subjected to identical conditions as in **a**. **e** Correlation of the maximum rotation rate shift with the time derivative of temperature for both the conventional PMF FOG and the four-tube *t*DNANF gyro over different temperature ranges. **f** Allan deviations of the static rotation rate measured by the four-tube *t*DNANF gyro at -40 °C, 25 °C, and 60 °C. The centre panel replicates the data presented in Fig. 4b.

To evaluate the thermal stability of the four-tube *t*DNANF gyro, a traditional PMF was quadrupolarly coiled with the same length and diameter to replace the four-tube *t*DNANF coil. The rotation rate shift evolution of the PMF gyro was similar to that of the four-tube *t*DNANF gyro during heating and cooling, but the PMF gyro was much more sensitive to transient temperature changes. Fig. 5d shows the PMF gyro's maximum rotation rate shift under the same conditions as Fig. 5a. The solid-core PMF gyro exhibited a larger rotation rate shift than the air-core four-tube *t*DNANF gyro. For example, under a temperature change rate of 1 °C per minute, the maximum rotation rate shifts of the PMF and four-tube *t*DNANF gyros were 0.071 deg h$^{-1}$ and 0.0079 deg h$^{-1}$, respectively.

The thermal sensitivities of both gyros were determined by plotting the maximum rotation rate shifts as a function of the applied temperature change rate (Figure 5e). The shifts varied linearly with the temperature change rate. Linear fitting revealed that the thermal sensitivity of the PMF gyro was approximately 0.0641/60 = 1.07×10$^{-3}$ deg °C$^{-1}$, while that of the four-tube *t*DNANF gyro was 0.0069/60 = 1.15×10$^{-4}$ deg °C$^{-1}$, 0.0060/60 = 1.00×10$^{-4}$ deg °C$^{-1}$, and 0.0094/60 =

$1.57 \times 10^{-4}$ deg °C$^{-1}$ for temperatures ranging from -40 to 60 °C, 20 to 60 °C, and -40 to 20 °C, respectively. This indicates that the four-tube $t$DNANF gyro is 9.24 (-40 to 60 °C), 10.68 (20 to 60 °C) and 6.82 (-40 to 20 °C) times less sensitive to temperature variation over different temperature ranges than the PMF gyro under the same coil size and thermal shock conditions.

The difference in the measured thermal sensitivities suggests that the thermal attributes of the materials used in the fibre coating and coil glue play a critical role. The key parameters of these materials, such as their viscoelastic properties and Young's modulus, change markedly with temperature.

In addition to the thermal excursion test, static measurements of the rotation rate for the four-tube $t$DNANF gyro at different temperatures and their Allan deviations are also presented in Fig. 5f. The long-term bias instabilities were 0.0064 deg h$^{-1}$ at 60 °C, 0.0014 deg h$^{-1}$ at 25 °C, and 0.0028 deg h$^{-1}$ at -40 °C. The slight discrepancy may arise from the thermally induced variations in the properties of the fibre coating or curing glue.

Table 1 compares the performances of representative air-core FOGs. The proposed four-tube $t$DNANF gyro outperforms all other air-core FOGs in terms of the ARW at high power, BS drift over 100 s, BI drift over a long enough period (8500 s), and thermal sensitivity reduction compared to a PMF gyro of the same coil size. Our work demonstrated a dramatic 35-fold decrease in BI drift compared with any previous report. To the best of our knowledge, this is the first case of navigation-grade air-core FOG.

## Discussion

The use of an air-core fibre in FOG is expected to significantly reduce the extraneous phase drift and noise associated with the Kerr effect, Faraday effect, thermal effects, and radiation damage in sensing fibres[16,12]. These improvements could lead to higher long-term stability, lower noise, and better cost, size, weight, and power (C-SWaP) performance. However, early efforts faced challenges such as severe backscattering and poor spatial/polarization mode purity in PBG-HCFs. Although the introduction of shunt structures in cladding to achieve high spatial mode purity in PBG-HCFs was proposed a decade ago[46], the stringent requirements on fibre bending and orientation due to the small index mismatch between higher-order core modes and shunt cladding modes complicate the use of such fibres in a quadrupolar-wound IFOG.

Conversely, the strategy of degenerating polarization states of the core fundamental modes by varying the glass membrane thickness in the core surround with a fourfold rotational symmetry shape has proven effective in generic HCFs[41-43,46] and has been adopted in this work. Another advantage of this core surround design is that it aligns the principal axes of various wavelengths in the same direction, which is necessary for IFOG applications.

The inherent drawback of PBG-HCFs is that a great amount of modal field overlaps with the air/glass interfaces[49], which causes severe backscattering and interpolarization crosstalk, fundamentally undermining the Sagnac interferometric effect in an FOG. The long-standing stagnation of the low-loss PBG-HCF development can also be attributed to this issue[50].

The advent of ARF, which features broader bandwidth, better higher-order-mode suppression, and a greater laser damage threshold than PBG-HCF, has spurred numerous new application scenarios of HCF technology in communications[51], quantum state transmission[52], and high-power laser delivery[53]. In the field of FOG, decades of technique refinements have converged on the minimum-scheme-based IFOG, both in electronic and photonic aspects. Therefore, adhering to this well-established scheme and simply extending it, for example, through direct fibre-to-chip coupling with no redundant optics, may better advance the ultimate performance and promote widespread technology adoption. Recent attempts at ARF-based RFOG[31,32] cannot leverage these advantages.

**Table 1 | Representative air-core FOGs and criteria of navigation-grade FOGs.**

| Type of fibre & gyro system | Coil length/diameter (m) | ARW at high power (deg h$^{-1/2}$) | Bias stability (deg h$^{-1}$) | Integration time (s) | Thermal stability | Ref. |
|---|---|---|---|---|---|---|
| 7-cell PBG-IFOG | 235/0.082 | --- | 2 (*drift*) | / | 6.5 | [15] |
| 7-cell PBG-RFOG | 50 (*Finesse* = 7)/--- | 0.96 | 20 | 10 | --- | [47] |
| 7-cell PMPBG-IFOG | 250/0.05 | 0.048 | 0.51 | 50 | --- | [48] |
| Kagome-RFOG | 5.6 (*Finesse* = 58.2)/0.13 | 0.04 | 0.15 | 200 | --- | [31] |
| 6-tube NANF-RFOG | 136 (*Finesse* = 14.5)/0.114 | 0.09 | 0.05 | 3600-36000 | --- | [32] |
| | | | 0.3 | 100 | | |
| 4-tube *t*DNANF-IFOG | 469/0.12 | 0.0038 $^a$ | 0.0014 @ 25 °C | 8500 | 9.24/10.68/6.82 $^b$ | This work |
| | | | 0.023 $^a$ | 100 | | |
| Navigation-grade criteria | --- | 0.017-0.0017 | 0.1-0.01 | 100 | --- | |

$^a$ the values measured at -40 °C, 25 °C, and 60 °C are the same, see Fig. 5f.
$^b$ the ratios between the thermal sensitivities of the solid-core PMF gyro and the air-core four-tube *t*DNANF gyro differ in the temperature ranges of -40 to 60 °C, 20 to 60 °C, and -40 to 20 °C, see Fig. 5e.

In summary, the 469 m four-tube $t$DNANF coil-based IFOG demonstrated in this work represents a significant step in the pursuit of high-performance air-core FOGs. With an ARW of 0.0038 deg h$^{-1/2}$ and a BI drift over 8500 s of 0.0014 deg h$^{-1}$, this gyro showcases navigation-grade performance above one order of magnitude better than the previous record[32]. Notably, an ~10-fold reduction in thermal sensitivity compared to that of a conventional PMF gyro using a coil of the same length has been experimentally validated, updating previous predictions on the thermal stability gain of air-core FOGs[16-18]. The thermal stability is expected to further improve by optimizing the thermal and mechanical properties of the filling glue in the fibre coils[54].

Based on our analysis of fibre optical properties, the next goal is to further refine the phase birefringence and principal axis offset, which appear feasible within the context of our ARF design, potentially making air-core IFOGs a real contender to their solid-core fibre counterparts.

## Methods
### Finite element method simulation
The effective index, vector field distribution, and confinement loss of a polarization mode were simulated using commercial finite-element solvers, such as COMSOL Multiphysics, with an optimized mesh size and a perfectly matched layer. The geometrical parameters were extracted from SEM images of the fibre, with some adjustments made within the range of uncertainties. Dielectrics were modelled using the Sellmeier equation for silica and considering $n = 1$ for air. To assess the macrobend loss of a fibre, the refractive index distribution was conformally mapped to $n_b = n_s \cdot e^{x/R_b}$, where $n_b$ is the refractive index distribution of the bent fibre, $n_s$ is that of the straight fibre, $R_b$ is the bend radius, and the fibre is bent towards the negative $x$-axis.

### Fibre design and fabrication
To date, all reported fabricated NANFs and DNANFs adopt five or more tubular units, where each outer tube encloses a mid-sized tube and a smaller tube. These tubular units are radially oriented around a central core without touching each other, resulting in several void regions behind the gaps. Four-tube DNANFs are seldom considered because the corresponding four void regions are equivalent in size to the core area, promoting unwanted coupling of the fundamental core mode with the cladding air mode. To mitigate this effect, the four outer tubes are truncated into oversized semicircle shapes, pronouncedly diminishing the area of the voids behind the gaps. The fabrication of these advanced $t$DNANFs entails precision cutting of the full tubes at a prescribed angle by high-power lasers. Subsequently, these truncated tubes are aligned and assembled with the inner middle and smaller tubes, resulting in a structured assembly. The entire arrangement undergoes thermal drawing to form a preform, which is further processed into the final fibre. During the fibre drawing process, differential pressurization is strategically applied to four zones—the core region, the truncated outer tubes, the middle-sized tubes, and the smallest tubes—to ensure the integrity and performance of the fibre.

### Loss measurement
Spectral attenuation was measured by the cut-back method from 510 to 10 m for the four-tube $t$DNANF. A supercontinuum (SC-5, YSL, 470 nm to 2400 nm) source was stably butt-coupled to the fibre (inside a fibre splicer), which was looped on a bobbin with the circumstance of 1 m. The output of the fibre was connected to an optical spectral analyser (AQ6370D, Yokogawa, 600 nm to 1700 nm) through a magnetic clamp bare fibre adaptor. Multiple cleavages of the fibre end revealed little variation in the recorded spectra.

Macrobend loss measurements were conducted by comparing the transmission spectrum of a bent fibre with that of a quasistraight fibre with a loop radius of 50 cm.

### Symmetrical quadrupolar fibre winding
The four-tube $t$DNANF was wound from the centre, alternating layers from each half-length to position symmetrical segments in proximity. The quadrupolar winding method, where the fibre layer order is reversed pair by pair, can greatly mitigate the Shupe effect of the entire coil. To achieve optimal results, the number of fibre layers should be a multiple of four, with each layer containing the same number of turns. During fibre winding, on-line transmission loss was monitored to adjust the tension in real-time, thus minimizing damage to the fibre.

### Spatial mode purity measurement
The modal purity of a fibre was analysed using the S$^2$ imaging method[45]. A tunable laser (Santec TSL-550A, 1490 nm to 1640 nm) transported light into the fibre under test, and an infrared CCD (WiDy Sens 320, NIT) camera determined the spatial distribution of the transmitted light as the wavelength swept across a range of 2.5 nm with a spacing of 1 pm. Subsequently, the multipath interference (MPI) at the fibre output, defined as the power of a higher-order mode relative to the fundamental mode, was calculated by applying Fourier transformation and then integrating the optical spectrum at each pixel in the cross section.

### Shupe constant measurement
The Shupe constant was measured using an all-fibre Mach–Zehnder interferometer (see Supplementary Material S1.2). A 400 kHz linewidth tunable laser (Santec TSL-550A, @1550 nm) was employed. In the signal arm of the interferometer, the fibre under test (FUT) was coiled with a diameter of 6 cm and placed in a homemade thermal chamber along with two thermometers (with a resolution of 0.1 °C). Both ends of the FUT were spliced with SSMF pigtails and incorporated into the interferometer. The length difference between the SSMFs in the signal and the reference arms was less than 2 cm to minimize unwanted phase drift caused by SSMFs. After a 3× 3 fibre optic coupler, the interferograms were acquired in real time by three InGaAs photodetectors (Thorlabs, PDA015C2) connected to a digital data acquisition card and then used to derive the optical phase change accumulated when light passes through the FUT .

**PER and principal axis angle measurements**
The polarization properties of our ARFs were measured by a crossed-polarizer setup, which consisted of a supercontinuum source, two calcite polarizers, two achromatic half-wave plates (HWP1/HWP2), and an optical spectrum analyser (see Supplementary Material S1.3). The fibre under test was coiled and placed in a thermal chamber. The input and the output polarization states were tuned by rotating HWP1 and HWP2, respectively. Since the interpolarization coupling in ARFs is weak, a coiled ARF can be regarded as a wave plate with a wavelength-dependent principal axis angle ($\theta_A$) and phase retardance ($\phi$) (see details in Supplementary Material S2.1). Over a modestly wide wavelength range (e.g., 1525-1565 nm in this study), the intensities transmitted through a crossed-polarizer setup can be approximately expressed as[36]

$$\begin{cases} I_{0°/0°}(\lambda) = 1 - \frac{1}{2}\sin^2(2\theta_A(\lambda)) \cdot (1 - \Delta \cdot \cos(\phi(\lambda))) \\ I_{0°/90°}(\lambda) = \frac{1}{2}\sin^2(2\theta_A(\lambda)) \cdot (1 - \Delta \cdot \cos(\phi(\lambda))) \\ I_{45°/45°}(\lambda) = 1 - \frac{1}{2}\cos^2(2\theta_A(\lambda)) \cdot (1 - \Delta \cdot \cos(\phi(\lambda))) \\ I_{45°/135°}(\lambda) = \frac{1}{2}\cos^2(2\theta_A(\lambda)) \cdot (1 - \Delta \cdot \cos(\phi(\lambda))) \end{cases}. \quad (2)$$

Here, the subscripts 0°/0°, 0°/90°, 45°/45°, and 45°/135° represent the input/output polarization angles of measurement, and an empirical less-than-one factor $\Delta$ accounts for the lossy interpolarization coupling between sequential segments along the fibre. From the measured spectra, the principal axis angle ($\theta_A$) can be derived from Eq. (2) using the same method as in Ref. 36 (the results of which are presented in Fig. 3e). Based on the measured $\theta_A(\lambda)$, a broadband PER (integrated from $\lambda_1$ to $\lambda_2$) can be assessed by the following expression when the phase retardance $\phi$ varies rapidly with wavelength considering a fibre length of hundreds of metres (the results of which are presented in Fig. 3f):

$$PER(dB) \approx 10 \times \log \frac{\int_{\lambda_1}^{\lambda_2} \left[1 - \frac{1}{2}\sin^2(2\theta_A(\lambda))\right] d\lambda}{\int_{\lambda_1}^{\lambda_2} \left[\frac{1}{2}\sin^2(2\theta_A(\lambda))\right] d\lambda}. \quad (3)$$

**Fibre-to-chip direct coupling**
Direct coupling assembly is used to prevent microstructure collapse of the four-tube *t*DNANF due to fusion. Two sets of microlenses are introduced between the pigtails of the four-tube *t*DNANF coil and the bare Y waveguide chip to collimate and converge the light, facilitating low-loss coupling. Additionally, a reflective microprism is placed between one set of microlenses. The collimated beam emitted from the microlenses is reflected twice, ensuring that the parallel light beams transmitted through the two pairs of microlenses are spatially separated. This separation serves two purposes: (1) it increases the transverse distance to prevent crosstalk between the two light beams from the Y waveguide, and (2) it provides sufficient space to facilitate direct coupling operations of the four-tube *t*DNANF.

**FOG static performance measurements**
To measure the FOG static performance, the four-tube *t*DNANF gyro was mounted on a stable pier at room temperature (25 °C) with an input rotation rate of only a fraction of the Earth's rotation rate, approximately 9.47 deg h$^{-1}$. At such a low and stable input rate, we can ascribe any drift or variation to random noise (ARW) over short integration times or bias stability over longer integration times. The random measurement signal of the gyro, $\Omega(t)$, can be averaged over an integration time $\tau$, yielding a series of averaged values $\overline{\Omega_k}(\tau)$. The Allan deviation, ADEV($\tau$), is defined as the square root of half the mean value of the square of the difference between two successive averaged values:

$$ADEV(\tau) = \sqrt{\frac{1}{2}\left\langle (\overline{\Omega_{k+1}}(\tau) - \overline{\Omega_k}(\tau))^2 \right\rangle}. \quad (4)$$

The Allan deviation indicates that the rate output variation of the gyro (rate uncertainty) is a function of the integration time. The ARW can be obtained from the Allan deviation value at an integration time of 1 second divided by 60. The bias stability at an arbitrary integration time is equal to the Allan deviation at the corresponding integration time. Bias instability appears on the Allan deviation plot as a flat region around the minimum.

**Thermal excursion measurements**
The gyro was stably placed on a vibration isolation base inside a temperature chamber (GWS, WKZT1-50C). The thermally induced rotation rate shift (thermal stability) was measured at various temperature change rates of 0.2, 0.5, 1, 2, and 5 °C per minute across different temperature ranges: -40 to 60 °C, 20 to 60 °C, and -40 to 20 °C.

**Data availability**
Relevant data supporting the key findings of this study are available within the article and the Supplementary Information file. All the raw data generated during the current study are available from the corresponding authors upon request.

## Acknowledgements

This research was supported by the National Natural Science Foundation of China (62075083, 12074349, 62222506, U21A20506, 62105122); the Basic and Applied Basic Research Foundation of Guangdong Province (2021B1515020030, 2021A1515011646, 2022A1515110218); and the Guangzhou Science and Technology Program (202201010460).

## Author contributions

The concept of this work was conceived by W. D. and M. L.. The *t*DNANF and NANF were designed and fabricated by S. G. and Y. W. The numerical simulation was performed by H. C. with the assistance of D. W., Y. S., and Y. W. The optical characterization of the *t*DNANF and NANF was implemented by Q. H. and H. W. with the assistance of Y. S. and W. D.. The quadrupolar fibre winding as well as the gyro test were carried out by M. L., F. H., M. Y., W. L., and X. Z.. The fibre-to-chip direct coupling and the temperature sensitivity measurements were carried out by M. L., and F. H.. All authors participated in the writing of the manuscript. The project was under the supervision of W. D., X. Z., and Y. W..

## Competing interests

The authors declare no competing interests.

## Additional information

**Supplementary information** The online version contains supplementary material available for this paper.

**Correspondence** and requests for materials should be addressed to Wei Ding, Xiaoming Zhao, or Yingying Wang.

# Navigation-grade interferometric air-core antiresonant fibre optic gyroscope with enhanced thermal stability

**Maochun Li[1,†], Shoufei Gao[2,3,5,†], Yizhi Sun[2,3,5,†], Xiaoming Zhao[1,\*], Wei Luo[1], Qingbo Hu[2,3], Hao Chen[2,3], Helin Wu[2,3], Fei Hui[1], Yingying Wang[2,3,5\*], Miao Yan[1,4], and Wei Ding[2,3,5,6\*]**

[1]Tianjin Key Laboratory of Quantum Precision Measurement Technology, Tianjin Navigation Instruments Research Institute, Tianjin 300131, China

[2]Guangdong Provincial Key Laboratory of Optical Fiber Sensing and Communication, Institute of Photonics Technology, Jinan University, Guangzhou 510632, China

[3]College of Physics & Optoelectronic Engineering, Jinan University, Guangzhou 510632, China

[4]School of Mechanical Engineering, Nanjing University of Science and Technology, Nanjing 210094, China

[5]Linfiber Technology (Nantong) Co., Ltd. Jiangsu 226010, China

[6]Peng Cheng Laboratory, Shenzhen 518055, China

[†]These authors contributed equally to this work.

Corresponding authors: *dingwei@jnu.edu.cn, *tjhhyq@yeah.net, *wangyy@jnu.edu.cn

# Contents

# S1. Experimental setups for fibre characterization

## S1.1 Higher-Order Modes

The relative amount of power carried by higher-order modes (HOMs) in NANFs is measured using the spatially and spectrally ($S^2$) imaging technique[1,2]. As depicted in Fig. S1, the setup comprises a tunable laser source (TLS, Santec TSL-550A, 1490 nm-1640 nm) and a CCD camera (WiDy Sens 320, NIT). The fibre under test (FUT) is a 510 m four-tube $t$DNANF used for the FOG. Light in the $t$DNANF is launched from a standard single-mode fibre (SSMF) by end-fire coupling. As the wavelength sweeps across a range of 2.5 nm with a step of 1 pm, the CCD camera records the spatial distributions of the transmitted light concurrently. Subsequently, the multipath interference (MPI) value, defined as the power ratio of the relevant modes at the output, can be calculated after Fourier analysis.

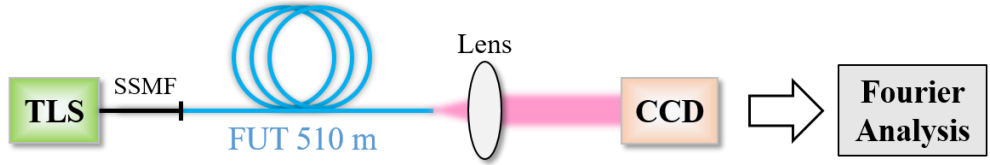

**Fig. S1.** Setup for $S^2$-imaging measurements.

## S1.2 Shupe Constant

The thermal sensitivity of the accumulated phase ($\varphi$) of light propagating through a fibre, referred to as the Shupe constant ($S$), can be expressed as[3]

$$S = \frac{1}{\varphi}\frac{d\varphi}{dT} = \frac{1}{L}\frac{dL}{dT} + \frac{1}{n_{eff}}\frac{dn_{eff}}{dT}, \tag{S1}$$

where $n_{eff}$ is the effective refractive index, $\lambda$ is the wavelength in vacuum, $L$ is the fibre length, and $T$ is the temperature.

Figure S2 shows the setup of the fibre Mach–Zehnder interferometer used in this work. A narrow linewidth laser (TLS, Santec TSL-550A, @1550 nm) with high frequency stability is employed. In the signal arm, the FUT is coiled with a radius of 6 cm and placed in a homemade thermal chamber alongside two thermometers (at a resolution of 0.1 °C). Both ends of the FUT are spliced to a piece of SSMF. The length difference between the SSMFs in the signal and the reference arms was less than 2 cm to minimize the environmental temperature-induced phase drift. After a 3×3 fibre optic coupler, the interferograms are acquired in real time by three InGaAs photodetectors (Thorlabs, PDA015C2) connected to a digital data acquisition card. Then, the thermally induced optical phase change over the FUT can be retrieved.

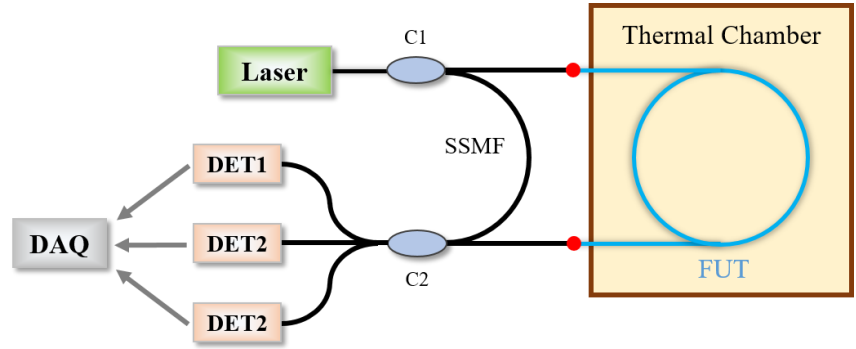

**Fig. S2.** Setup for Shupe constant measurement. C1, 2×2 fibre optic coupler (50:50); C2, 3×3 fibre optic coupler (33:33:33); Red points, fusion splices between the SSMF and *t*DNANF; DET1-3, photodetectors; DAQ, data acquisition card.

### S1.3 Linear Polarization Purity

To measure the polarization property of our *t*DNANF, a polarization measurement setup was built using an SC, two calcite polarizers (P1/P2), two achromatic half-wave plates (HWP1/HWP2), and an OSA (Fig. S3). The FUT is coiled with a radius of 6 cm and placed in the thermal chamber. The input and output polarization states of the FUT are tuned by rotating HWP1 and HWP2, respectively.

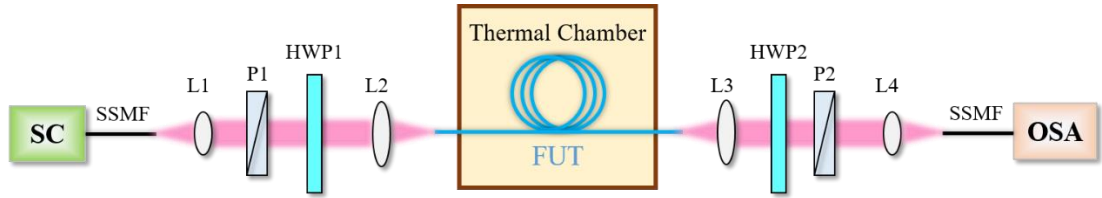

**Fig. S3.** Setup for linear polarization measurements. SC, supercontinuum source; L1-L4, lens; P1/P2, polarizers; HWP1/HWP2, half-wave plates; OSA, optical spectrum analyser.

## S2. Modelling and measurement of polarization properties

### S2.1 Modelling

To facilitate the analysis of the polarization measurements, it is imperative to model the birefringence properties of an AR-HCF, which are primarily dictated by two factors. The first is intrinsic and related to the microstructure of an AR-HCF, including the ellipticity of the core and the nonuniformity of the membrane thickness[4]. The second factor is extrinsic, including macro/microbending and twisting. As a result, an AR-HCF can be treated as a cascade of wave plates (see Fig. S4), and each wave plate possesses a wavelength-dependent principal axis angle and phase retardance. When an AR-HCF is placed stably and under static temperature, the entire AR-HCF can be equivalently represented as a wavelength-dependent wave plate.

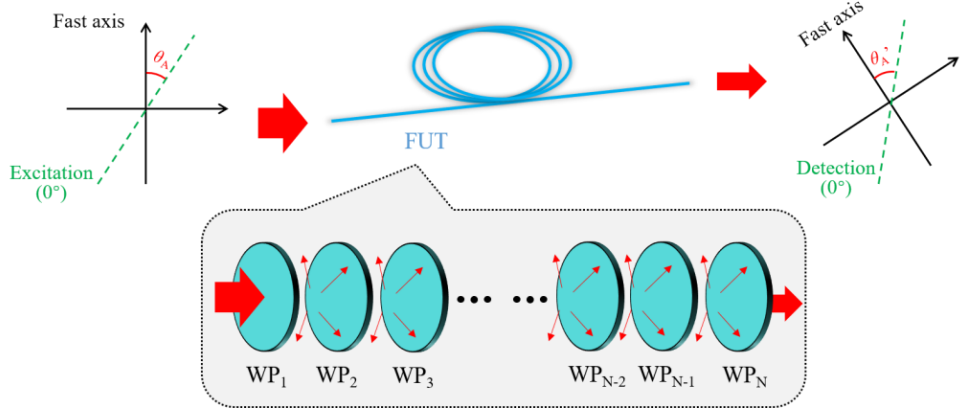

**Fig. S4.** Schematic of the AR-HCF as a birefringent device for crossed-polarizer transmission. WP: ideal wave plate.

During measurement, (0°/0°) refers to the situation where the excitation and detection polarization angles are aligned to the input and the output principal axes, respectively, by rotating the two HWPs. As shown in Fig. S4, $\theta_A$ is the angle between the excitation polarization and the fast axis, and $\theta'_A$ is the angle between the detection polarization and the fast axis. Note that both of these angles are wavelength dependent and can only be zero at certain wavelengths. By rotating the two HWPs, the normalized transmission spectra can be expressed as

$$\begin{cases} I_{0°/0°}(\lambda) = \cos[\theta_A(\lambda) - \theta'_A(\lambda)]^2 - \frac{1}{2}\sin(2\theta_A(\lambda)) \cdot \sin(2\theta'_A(\lambda)) \cdot (1 - \Delta \cdot \cos(\phi(\lambda))) \\ I_{0°/90°}(\lambda) = \sin[\theta_A(\lambda) - \theta'_A(\lambda)]^2 + \frac{1}{2}\sin(2\theta_A(\lambda)) \cdot \sin(2\theta'_A(\lambda)) \cdot (1 - \Delta \cdot \cos(\phi(\lambda))) \\ I_{45°/45°}(\lambda) = \cos[\theta_A(\lambda) - \theta'_A(\lambda)]^2 - \frac{1}{2}\cos(2\theta_A(\lambda)) \cdot \cos(2\theta'_A(\lambda)) \cdot (1 - \Delta \cdot \cos(\phi(\lambda))) \\ I_{45°/135°}(\lambda) = \sin[\theta_A(\lambda) - \theta'_A(\lambda)]^2 + \frac{1}{2}\cos(2\theta_A(\lambda)) \cdot \cos(2\theta'_A(\lambda)) \cdot (1 - \Delta \cdot \cos(\phi(\lambda))) \end{cases} \quad (S2)$$

where $\phi(\lambda)$ is the wavelength-dependent phase retardance accumulated between the two principal axes, whose spectrum can be used to determine the group birefringence. $0 < \Delta < 1$ represents an empirical correction factor to account for the lossy coupling between adjacent fibre segments or the 'wave plates' outlined in Fig. S4. For the excitation and detection polarization angles, $\theta_A(\lambda)$ does not need to equal $\theta'_A(\lambda)$ across the whole transmission wavelength region of the AR-HCF, which spans hundreds of nanometers. This inequality causes difficulties in calculating the principal axis angles by using Eq. (S2).

In our measurement, we first adjust the two HWPs to obtain the minimum transmission power at a single wavelength ($\lambda_0$, i.e., 1545 nm in this work), which corresponds to a crossed-polarizer measurement (0°/90°) at $\theta_A(\lambda_0) = \theta'_A(\lambda_0) = 0$. In proximity to $\lambda_0$ (1525 - 1565 nm in this work), we hypothesize $\theta_A(\lambda) \approx \theta'_A(\lambda)$ and therefore simplify Eq. (S2) to a form similar to the expression in Ref. [5]:

$$\begin{cases} I_{0°/0°}(\lambda) = 1 - \frac{1}{2}\sin^2(2\theta_A(\lambda)) \cdot (1 - \Delta \cdot \cos(\phi(\lambda))) \\ I_{0°/90°}(\lambda) = \frac{1}{2}\sin^2(2\theta_A(\lambda)) \cdot (1 - \Delta \cdot \cos(\phi(\lambda))) \\ I_{45°/45°}(\lambda) = 1 - \frac{1}{2}\cos^2(2\theta_A(\lambda)) \cdot (1 - \Delta \cdot \cos(\phi(\lambda))) \\ I_{45°/135°}(\lambda) = \frac{1}{2}\cos^2(2\theta_A(\lambda)) \cdot (1 - \Delta \cdot \cos(\phi(\lambda))) \end{cases} \quad (S3)$$

After acquiring the transmission spectra in the configurations of 0°/0°, 0°/90°, 45°/45°, and 45°/135°, the wavelength dependences of $\theta_A(\lambda)$ [$\approx \theta'_A(\lambda)$] can be derived.

Based on the measured $\theta_A(\lambda)$, the broadband PER can be estimated by the following calculation when the input power spectrum is assumed to be uniform from $\lambda_1$ to $\lambda_2$:

$$\begin{aligned} PER(dB) &= 10 \times \log \frac{\int_{\lambda_1}^{\lambda_2} I_{0°/0°}(\lambda) d\lambda}{\int_{\lambda_1}^{\lambda_2} I_{0°/90°}(\lambda) d\lambda} \\ &= 10 \times \log \frac{\int_{\lambda_1}^{\lambda_2} \left[1 - \frac{1}{2}\sin^2(2\theta_A(\lambda)) \cdot (1 - \Delta\cos(\phi(\lambda)))\right] d\lambda}{\int_{\lambda_1}^{\lambda_2} \left[\frac{1}{2}\sin^2(2\theta_A(\lambda)) \cdot (1 - \Delta\cos(\phi(\lambda)))\right] d\lambda}, \end{aligned} \quad (S4)$$

where $\lambda_1$ is 1525 nm and $\lambda_2$ is 1565 nm. Considering group birefringence at the level of $10^{-6}$ to $10^{-5}$, the phase retardance $\phi(\lambda)$ can vary by tens of $2\pi$ from 1525 nm to 1565 nm with a fibre length of hundreds of metres. Therefore, the integrals in Eq. S5 can be simplified as

$$PER(dB) \approx 10 \times \log \frac{\int_{\lambda_1}^{\lambda_2} \left[1 - \frac{1}{2}\sin^2(2\theta_A(\lambda))\right] d\lambda}{\int_{\lambda_1}^{\lambda_2} \left[\frac{1}{2}\sin^2(2\theta_A(\lambda))\right] d\lambda}. \quad (S5)$$

### S2.2 Crossed-Polarizer Transmission Spectra

By utilizing the polarization measurement setup shown in Fig. S4, the transmission spectrum of the fibre can be obtained when the input/output linear polarization angles are rotated to 0°/0°, 0°/90°, and 45°/135°. Fig. S5(a) shows a typical result when a 3 m four-tube *t*DNANF is coiled with $R_b$ = 6 cm at room temperature. Across the broad wavelength range of the transmission window of this fibre, the minimum transmission is only observed at ~1550 nm, where $\theta_A = \theta'_A = 0$. Within the wavelength range near 1550 nm, the principal axis angle slightly changes, $\theta_A \approx \theta'_A$ [the orange region in Fig. S5(a)], and Eq. (S3) can be used to retrieve $\theta_A$. When the wavelength deviates far from 1550 nm [the green region in Fig. S5(a)], Eq. (S3) is no longer applicable because $\theta_A \neq \theta'_A$.

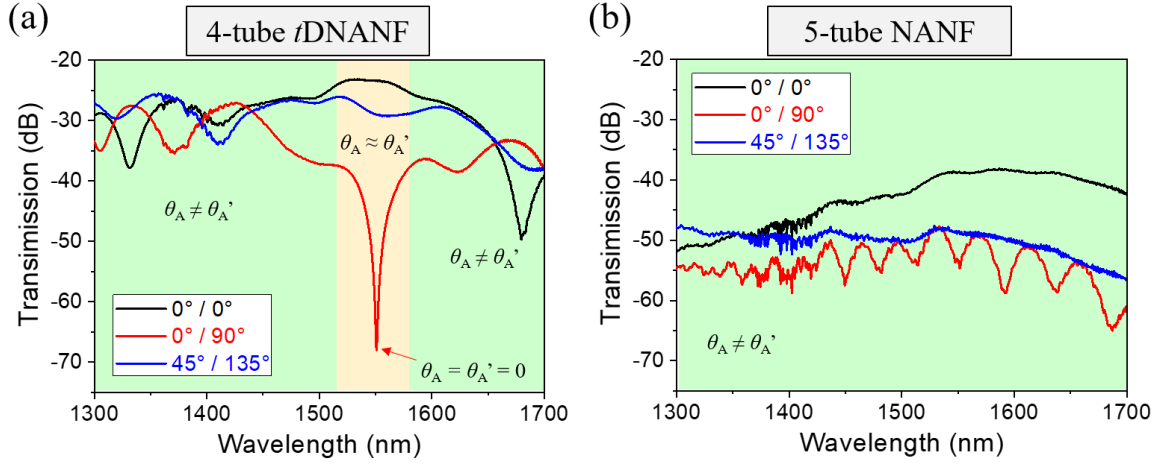

**Fig. S5.** Crossed-polarizer transmission spectra through 3 m (a) four-tube $t$DNANF and (b) five-tube NANF with $R_b = 6$ cm.

On the other hand, when considering a five-tube NANF (as mentioned in Fig. 2 of the main text), small-$\theta_A$ regions are rare, as depicted in Fig. S5(b). The cross-polarization transmission spectra imply a rapid variation in the principal axis angle with increasing wavelength. It seems that a $10^{-7}$ level of (low) birefringence cannot diminish the wavelength dependence of the principal axis angle. As a result, broadband PER will remain at a low level in this structure of ARFs.

**S2.3 Principal Axis Angle Variations in the Four-tube $t$DNANF**

Based on the crossed-polarizer transmission spectrum measurements and the model of Eq. (S3), our four-tube $t$DNANF is tested under different lengths and bend radii. As shown in Fig. S6, the measured principal axis offsets vary under different conditions, indicating that imperfect winding with twisting and stress can slightly influence the principal axis. Such imperfections are harder to eliminate in longer fibres. Nevertheless, in the wavelength region of interest (1525-1565 nm), the variation is relatively small, thus ensuring a broadband PER of ~ 20 dB. This result can be attributed to the modest but not small birefringence (at the $10^{-6}$ level) provided by our four-tube $t$DNANF.

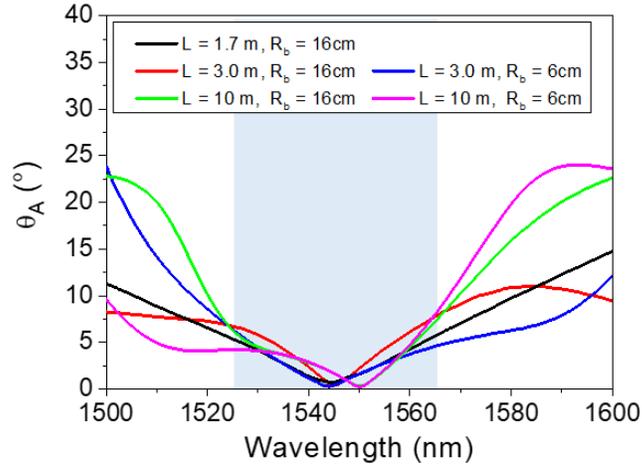

**Fig. S6.** Measured principal axis offsets relative to the orientations under different fibre lengths and bend radii at room temperature.

## S3. Manufacture and test of the four-tube *t*DNANF gyro

### S3.1 Symmetrical Quadrupolar Fibre Winding

During symmetrical quadrupolar fibre winding, on-line transmission loss is monitored to adjust the tension in time, thereby minimizing damage to the fibre. As shown in Fig. S7(a), an ASE source of 40 nm bandwidth and a power meter are connected to the two ends of the four-tube *t*DNANF rotating with fibre suppliers to provide for on-line transmission loss monitoring. The quadrupolar winding method, where the fibre layer order is reversed pair by pair (as shown in Fig. S7(b)), can greatly mitigate the Shupe effect of the whole coil.

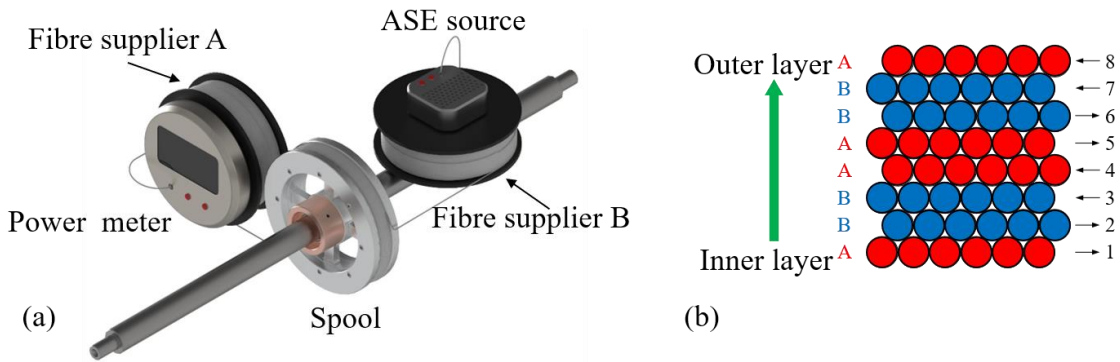

**Fig. S7.** Schematic of symmetrical quadrupolar fibre winding. (a) The online transmission loss monitoring setup. (b) Section view of symmetrical quadrupolar winding.

### S3.2 Fibre-to-Chip Direct Coupling

On the fibre-to-chip direct coupling platform shown in Fig. S8, the four-tube *t*DNANF coil ends are directly connected to the multifunction integrated optics chip (MIOC). This operating platform consists of a set of motion control systems, a power meter, an ER meter, and other components.

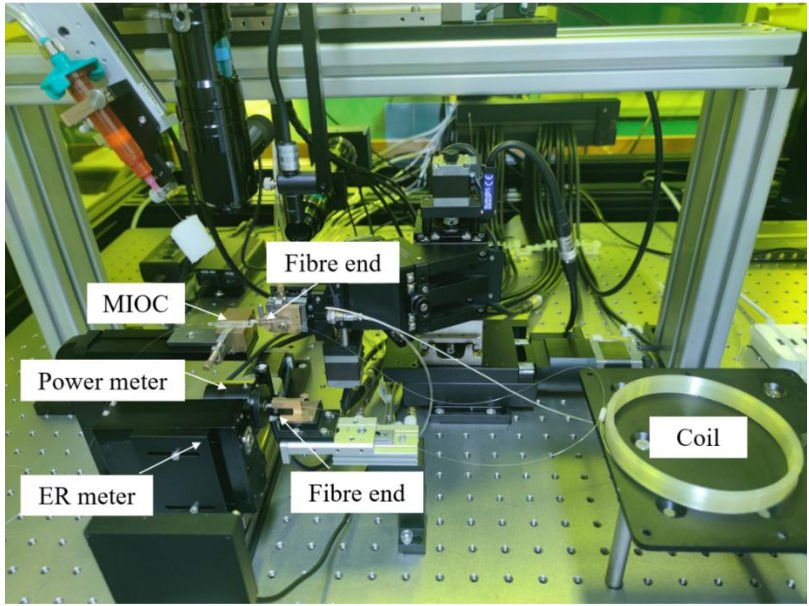

**Fig. S8.** Fibre-to-chip direct coupling assembly platform.

**S3.3 Overall Dimensions of the Four-tube *t*DNANF Gyro**

A photograph of the four-tube *t*DNANF gyro is shown in Fig. S9, which adopts an optic-electronic separation structure design. The photonic subsystem consists of a fibre coil and an MIOC, whose diameter and height are 156 mm and 20 mm, respectively. The electronic subsystem, with a size of 97.5×97.5×35 mm$^3$, consists of an ASE, a PMFC, a PD, and an electronics package.

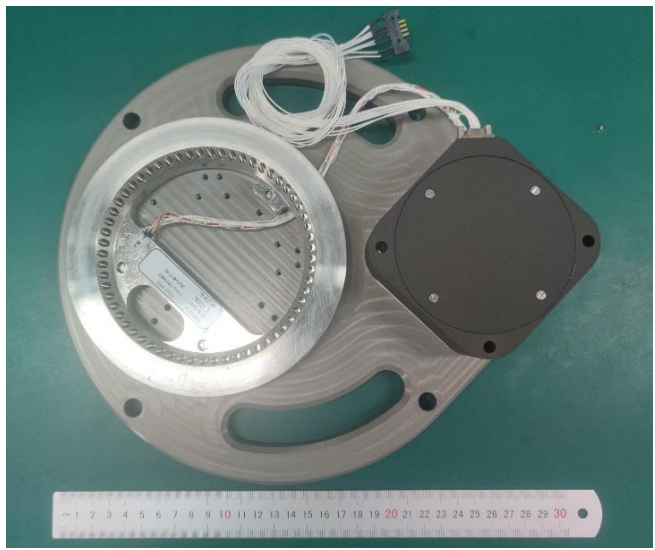

**Fig. S9.** Photograph of the four-tube *t*DNANF gyro.

**S3.4 FOG Static Performance Test**

The four-tube *t*DNANF gyro is mounted on a stable pier (as shown in Fig. S10) with an input rotation rate of only a fraction of the Earth's rate, and the output data of the gyro are stored in a computer. At such a low and stable input rate, any drift or variation can be attributed to random noise (ARW) over short integration times or bias stability/instability (BS/BI) over longer integration times.

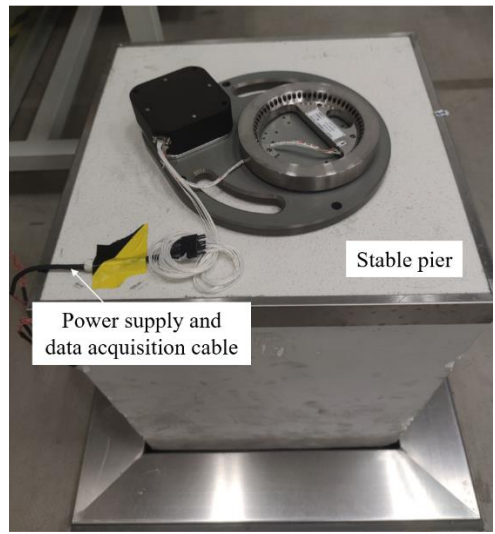

**Fig. S10.** Static performance test setup.

**S3.5 FOG Thermal Excursion Measurement**

The four-tube *t*DNANF gyro is stably placed on a vibration isolation base inside a temperature chamber (as shown in Fig. S11). The thermally induced rotation shift is measured at various temperature change rates across different temperature ranges.

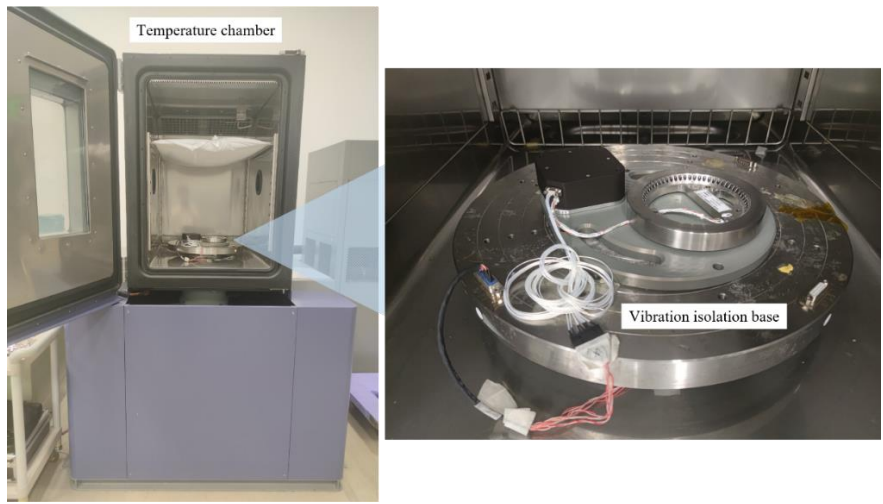

**Fig. S11.** Thermal excursion measurement setup.